\documentclass[english]{article}
\usepackage[T1]{fontenc}
\usepackage[latin9]{inputenc}
\usepackage{geometry,times}
\usepackage{color}
\usepackage{amsmath}
\usepackage{amssymb}
\usepackage{stackrel}
\usepackage{graphicx}
\usepackage{esint}
\usepackage{epstopdf}
\usepackage{epsfig}

\makeatletter

\providecommand{\tabularnewline}{\\}

\usepackage{babel}
\usepackage{cite}

\usepackage{babel}

\makeatother

\usepackage{babel}
\begin{document}

\title{Orthogonal-state-based and semi-quantum protocols for quantum private
comparison in noisy environment}

\author{Kishore Thapliyal$^{a,}$\thanks{Email: tkishore36@yahoo.com}, Rishi
Dutt Sharma$^{b,c,}$\thanks{Email: rishi.nitp@gmail.com}, and Anirban
Pathak$^{a,}$\thanks{Email: anirban.pathak@gmail.com}\\
 $^{a}$Jaypee Institute of Information Technology, A-10, Sector-62,
Noida, UP-201307, India\\
 $^{b}$National Institute Technology Patna, Ashok Rajhpath, Patna,
Bihar 800005, India\\
$^{c}$Department of Computer Science Engineering, Bennett University, Greater Noida, UP-201310, India}
\maketitle
\begin{abstract}
Private comparison is a primitive for many cryptographic tasks, and
recently several schemes for the quantum private comparison (QPC)
have been proposed, where two users can compare the equality of their
secrets with the help of a semi-honest third party (TP) without knowing
each other's secret and without disclosing the same to the TP. In
the exisiting schemes, secrecy is obtained by using conjugate coding,
and considering all participants as quantum users who can perform
measurement(s) and/or create states in basis other than computational
basis. In contrast, here we propose two new protocols for QPC, first
of which does not use conjugate coding (uses orthogonal states only)
and the second one allows the users other than TP to be classical
whose activities are restricted to either reflecting a quantum state
or measuring it in computational basis. Further, the performance of
the protocols is evaluated under various noise models. 
\end{abstract}
{\textbf{{Keywords:}} quantum private comparison, secure multiparty
computation, socialist millionaire problem, quantum cryptography,
noise models, quantum communication in noisy environment}

\section{Introduction\label{sec:Introduction}}

One of the most important branches of classical and quantum cryptography
is secure multi-party computation (SMC) (\cite{Yao-relates-all-task,SMC-sc-rep}
and references therein). SMC is a primitive for distributed computation.
It enables the distributed computing of correct output of a function
in a situation, where the inputs are given by a group of mutually
distrustful users. A SMC is required to be fair, and secure. Specifically,
it should not leak the secret inputs of the individual players. Efforts
have been made to achieve this in various ways, both classically \cite{Yao-relates-all-task,Boudot-equal-smc,sm2}
and quantum mechanically \cite{review-2014,yang-jphys-2009-step4-dishonesty-check-for-tp-should-be-taken-in-hwang,physscr_80_6_065002-with-correction,chi-type-state-ijtp-2012-liu.,hwang-qinp-2012,big-data}.
However, in those efforts \cite{chi-type-state-ijtp-2012-liu.,hwang-qinp-2012,big-data},
it has been assumed that some of the users follow the protocol honestly
(which implies that some of the users are semi-honest). Among the
variants of SMC schemes, a specific variant having particular importance
is ``socialist millionaire problem'' which was first introduced
by Yao \cite{Yao-relates-all-task} in 1982 as a computing task where
two millionaires (Alice and Bob) wish to know who is richer without
disclosing the amount of their wealth to each other. Subsequently,
Boudot et al. \cite{Boudot-equal-smc}, modified it to a task where
the millionaires are only interested to know whether they are equally
rich or not. Thus, they wish to compute a function $f(i_{A},j_{B}):\,f(i_{A},j_{B})=a\neq b=f(i_{A},i_{B})$,
where the subscripts $A$ and $B$ represent inputs from Alice and
Bob, respectively. Thus, $i_{A}$ and $i_{B}$ are essentially classical
information corresponding to the assets of Alice and Bob, and a scheme
for Boudot's version of the socialist millionaire problem is actually
a scheme for private comparison of equality of information. A lot
of work has been done on classical private comparison, but as the
security of all classical cryptographic schemes are conditional, it
can never lead to an unconditionally secure scheme of private comparison.
In contrast, we can achieve unconditional security in the quantum
world. This fact led to many proposals for private comparison of equality
of information using quantum resources \cite{review-2014,chi-type-state-ijtp-2012-liu.,hwang-ijtp-multiparty-2016,hwang-qinp-2012,Liu-j.optcom.2011-Wstate,multiparty-hwang-qinp-2015,noise-arvind-qinp-2015,noise-qinp-2015-collective-AD,physscr_80_6_065002-with-correction,SMC-sc-rep,sun-2015-qinp,yang-2013-qinp-comment-on-Hwang-2012,yang-jphys-2009-step4-dishonesty-check-for-tp-should-be-taken-in-hwang,zhang-ijtp-2015ent-swapping,zhang-qinp-2013-cryptanalysis-yang-2013,zhang-qinp2015-without-tp-cryptanalysis},
we may refer to all such protocols as protocols for ``quantum private
comparison (QPC)''. Before we proceed further, it would be apt to
note that QPC problem, the socialist millionaire problem and Tierce
problem \cite{Boudot-equal-smc} are equivalent, and in what follows
we will mostly refer to all such problems as QPC problem.

In the original definition of socialist millionaire problem, it was
a two-party computation task, but a pioneering work of Lo \cite{PhysRevA.56.1154-LO-97}
established that two-party secure QPC is not possible. This implies
that to implement a secure QPC, we must have a third party (TP), who
would assist the users to compare the equality of their secrets. Interestingly,
the TP may be semi-honest \cite{hwang-qinp-2012,chi-type-state-ijtp-2012-liu.},
semi-honest having an intelligent robot \cite{big-data}, dishonest
\cite{yang-2013-qinp-comment-on-Hwang-2012}, or almost-dishonest
\cite{hwang-ijtp-multiparty-2016}. Despite, the strong proof of Lo,
some efforts have been made to achieve QPC without TP \cite{without-TP1,without-tp2},
but they have been found to be insecure and/or unfair \cite{without-tp3}.
Thus, in what follows, we will concentrate on three party protocols
of QPC, where a TP helps Alice and Bob to compare the equality of
their information. Such protocols for QPC have been proposed using
different types of entangled states. For example, in Ref. \cite{chi-type-state-ijtp-2012-liu.},
a scheme for QPC has been proposed using $\chi$-type state, $W$
state was used in Ref. \cite{Liu-j.optcom.2011-Wstate}, Bell state
was used in \cite{hwang-qinp-2012,physscr_80_6_065002-with-correction,yang-2013-qinp-comment-on-Hwang-2012},
GHZ state was used in \cite{hwang-ijtp-multiparty-2016}. Further,
in Ref. \cite{QD}, a group theoretic structure of the protocols of
quantum dialogue was proposed using a large number of different types
of entangled states (e.g., $W,\,GHZ,\,{\rm cluster},\,Q_{4,}\,Q_{5,}$
and Brown states), and it was shown that the quantum dialogue scheme
proposed there can be converted to a protocol of the socialist millionaire
problem, which is equivalent to QPC. Thus, in \cite{QD}, several
options for realization of protocols for QPC were provided. Further,
in the similar line, in \cite{AQD}, a set of new options (e.g., 4-qubit
$\Omega$ state, 4-qubit cat state, etc.) for realization of QD, and
thus, QPC have been proposed.

It's already established that the schemes for QPC have useful applications
in private bidding and auctions, secret ballot elections, e-commerce,
etc. (\cite{review-2014} and references therein). Due to the fact
that a scheme for QPC has applications in many fields, many variants
of QPC have been studied in the recent past. For example, Huang et
al., have recently proposed a GHZ-state-based QPC scheme for $n$
users \cite{hwang-ijtp-multiparty-2016}. Huang et al.'s scheme considers
an almost-dishonest TP and allows him to compare the equality of the
secrets of a subset of users, too.

In what follows, we plan to propose two protocols for QPC in the line
of modified Tseng-Lin-Hwang (TLH) protocol, which was proposed in
its original form in 2012 \cite{hwang-qinp-2012}. In the original
TLH protocol a scheme for quantum private comparison was proposed
using Bell states, but almost immediately after the publication of
TLH scheme, Yang et al., \cite{yang-2013-qinp-comment-on-Hwang-2012}
had shown that there exist a security flaw in the original TLH scheme
and other similar schemes, which assume TP to be semi-honest. Yang
et al., \cite{yang-2013-qinp-comment-on-Hwang-2012} also proposed
a modified scheme which they claimed to be free from the limitations
of the previous protocols as their scheme considered a dishonest TP.
Subsequently, in an interesting work on cryptanalysis of Yang et al.'s
scheme, it was observed by Zhang et al., \cite{zhang-qinp-2013-cryptanalysis-yang-2013}
that Yang et al.'s scheme was also not completely secure. Keeping
these facts in mind, here we aim to propose two schemes of QPC that
are free from the attacks described earlier and are fundamentally
different from all the existing schemes of QPC as one of the proposed
schemes is semi-quantum in nature and the other one is a completely
orthogonal-state-based scheme. Here it would be apt to note that protocols
based on orthogonal states alone and semi-quantum protocols are foundationally
different from usual conjugate-coding based (BB84-type) protocols.
This is so because, in contrast to common perception developed by
BB84-type schemes, orthogonal-state-based schemes establish the fact
that unconditional security does not originate from the conjugate
coding (our inability to simultaneously measure a quantum state using
two or more MUBs). Further, semi-quantum schemes establish that all
parties involved in performing the cryptographic tasks do not require
to be quantum. A brief review of the existing orthogonal-state-based
and semi-quantum schemes for various quantum cryptographic task is
provided in the next section. The review clearly shows that to the
best of our knowledge, there does not exist any orthogonal-state-based
or semi-quantum scheme for QPC. This fact further sets the motivation
for designing such schemes for the first time.

In a practical situation, all communication schemes are affected by
the noise present in the communication channel. So it's extremely
important to know how QPC schemes behave in the presence of noise.
Recently, some efforts have been made to investigate the effect of
noise on QPC schemes. For example, the effects of Pauli noise \cite{noise-arvind-qinp-2015},
and collective amplitude damping \cite{noise-qinp-2015-collective-AD}
have been studied very recently. However, to the best of our knowledge,
no serious effort has yet been made to rigorously investigate the
effect of a complete set of possible noise models on the schemes of
QPC. This letter aims to do that for the protocols proposed in this
letter.

Remaining part of the letter is organized as follows. In Section \ref{sec:Protocols-for-quantum},
we have proposed two schemes of QPC that are fundamentally different
from the existing schemes of QPC. In Section \ref{sec:Analysis-of-security-efficiency},
the security and efficiency of the proposed schemes are analyzed.
The effect of various types of noise models on the proposed schemes
is investigated in Section \ref{sec:Effect-of-different-noise}.
Finally, the letter is concluded in Section \ref{sec:Conclusion}.

\section{Protocols for quantum private comparison\label{sec:Protocols-for-quantum}}

In this section, we introduce two protocols for QPC. The first one
is an orthogonal-state-based protocol, and the second one is a semi-quantum
protocol. Here, we assume that Alice and Bob wish to compare their
assets with the help of a TP.

\subsection{Orthogonal-state-based protocol for quantum private comparison}

In quantum cryptography, orthogonal-state-based protocols are of fundamental
interest as in contrast to BB84 type conjugate coding based protocol,
where the security arises from the conjugate coding or noncommutivity,
here (i.e., in case of orthogonal-state-based schemes), the security
arises from the duality (monogamy of entanglement) for single particle
(multipartite entangled) states. The first ever orthogonal-state-based
scheme of QKD was introduced in 1995 \cite{vaidman-goldenberg}, and
is now known as GV protocol. Since then several quantum cryptographic
schemes based on orthogonal states have been proposed for various
cryptographic tasks. For example, orthogonal-state-based schemes have
been proposed for QKA \cite{QKA-cs}, QSDC \cite{book,With-preeti,With-chitra-IJQI},
DSQC \cite{book,With-chitra-IJQI,dsqc-ent-swap}, CDSQC \cite{cdsqc},
etc. A few of these schemes have also been experimentally realized
\cite{GV-experiment}. However, to the best of our knowledge, no effort
has yet been made to design an orthogonal-state-based protocol for
quantum private comparison. In what follows, the same is proposed
in 8 steps, which are denoted by $({\rm OSB1,OSB2,\cdots OSB8)},$
where OSB stands for orthogonal-state-based. This notation is adopted
to distinguish steps of this protocol from the steps of our second
protocol, which is a semi-quantum scheme, and whose steps are denoted
by $({\rm SQ1,\cdots SQ9),}$ where ${\rm SQ}$ stands for semi-quantum.
The proposed orthogonal-state-based scheme works as follows: 
\begin{description}
\item [{OSB1:}] TP prepares $2N$-EPR pairs randomly chosen from the set
of four Bell states $\left\{ |\psi^{\pm}\rangle=\frac{|00\rangle\pm|11\rangle}{\sqrt{2}},\,|\phi^{\pm}\rangle=\frac{|01\rangle\pm|10\rangle}{\sqrt{2}}\right\} $.
TP prepares two quantum sequences, namely $S_{A}$ and $S_{B}$, which
contain all the first and second particles of the EPR pairs, respectively.
\\
 Thus, both $S_{A}$ and $S_{B}$, contain $2N$ qubits, and if a sequence
is transmitted via quantum channel, we would require to check half
of the travel qubits for eavesdropping. The same can be done by adding
$2N$ decoy qubits in each sequence. 
\item [{OSB2:}] TP prepares two copies of $|\psi^{+}\rangle^{N}$.
He uses the first (second) copy as a set of decoy qubits $D_{A}$
($D_{B}$) to be randomly inserted in the sequence $S_{A}\,(S_{B})$
to perform eavesdropping check using GV-subroutine (detail of GV-subroutine
can be found in \cite{With-preeti,With-chitra-IJQI,RDshrama2015}).
Specifically, TP randomly inserts $D_{A}$ in $S_{A}$ ($D_{B}$ in
$S_{B}$) to obtain a new enlarged sequence $S_{A}^{\star}$ $\left(S_{B}^{\star}\right)$.
Later, he sends the quantum sequence $S_{A}^{\star}$ to Alice and
$S_{B}^{\star}$ to Bob. 
\item [{OSB3:}] After receiving quantum sequences $S_{A/B}^{\star}$ Alice/Bob
perform GV subroutine for eavesdropping check with the help of TP.
They compute the error rate, if this error rate is more than a predetermined
allowed threshold value, then they abort the protocol and restarts
from OSB1 considering that an eavesdropper is present. Otherwise,
they proceed to the next step.\\
 As the eavesdropping is checked by measuring the decoy qubits, after
this step, Alice and Bob are left with $S{}_{A}$ and $S{}_{B}$,
respectively, because they remove the qubits measured for eavesdropping
check. 
\item [{OSB4:}] Qubits of $S{}_{A}$ and $S{}_{B}$ are now measured in the
computational basis by Alice and Bob, respectively. Both would individually
obtain $2N$-bits key strings of bit values $0$ and $1$ corresponding
to the measurement outcomes $|0\rangle$ and $|1\rangle$, respectively.\\
In half of the shared secret bits, they can check correlations to enhance security against disturbance attacks by Eve.
The reduced bit string of $N$-bits with Alice (Bob) is denoted as $K_{A}$ ($K_{B}$). 
\item [{OSB5:}] Alice and Bob prepare a shared key $K$$_{AB}$ of $N$-bits
using an orthogonal-state-based scheme of QKD \cite{vaidman-goldenberg}
or quantum key agreement (QKA) \cite{QKA-cs}. 
\item [{OSB6:}] Alice and Bob have the private information regarding their
assets which they wish to compare. Consider that the private information
of Alice and Bob is represented by the bit strings $M_{A}$ and $M_{B}$,
respectively. Alice encrypts her secret $M_{A}$ with her key string
$K_{A}$ and shared key $K$$_{AB}$ by using an exclusive-OR operation
to obtain $C_{A}^{i}=\left(M_{A}^{i}\oplus K_{A}^{i}\oplus K_{AB}^{i}\right)$.
Meanwhile, Bob also encrypts his message $M_{B}$ with keys $K_{B}$
and $K$$_{AB}$ by using the same operation to obtain $C_{B}^{i}=\left(M_{B}^{i}\oplus K_{B}^{i}\oplus K_{AB}^{i}\right)$.
Here, the superscript $i$ denotes the $i$th bit of the $N$-bits
string. Alice and Bob send the calculated strings $C_{A}$ and $C_{B}$
separately to TP via a public channel. 
\item [{OSB7:}] TP generates a classical bit string $C_{TP}$ of $N$-bits
corresponding to the choice of the initial Bell states in OSB1, such
that for the $i$th Bell state being $|\psi^{\pm}\rangle$ $\left(|\phi^{\pm}\rangle\right)$
he generates $i$th bit value in $C_{TP}$ as 0 (1). 
\item [{OSB8:}] TP computes $R$, which is now exclusive-OR result of $C_{A}$,
$C_{B}$, and $C_{TP}$ as $R=\left(C_{A}\oplus C_{B}\oplus C_{TP}\right)$.
The bit value 0 (1) of $R^{i}$ corresponds to the same (different)
values of $M_{A}^{i}$ and $M_{B}^{i}$. Thus, iff $R^{i}=0\,:\forall i\in\left\{ 1,\ldots,N\right\} $
both Alice and Bob have the same amount of assets. Checking values
of $R^{i}$, TP announces whether Alice and Bob have equal amount
of assets. 
\end{description}
Interestingly, following similar argument, the existing controlled
QD scheme proposed by some of the present authors in Ref. \cite{crypt-switch}
can also be modified to design a completely orthogonal-state-based
QPC protocol. However, we are not interested to elaborately discuss
another orthogonal-state-based scheme. Instead, here, we prefer to
propose another foundationally relevant scheme which can work with
limited quantum resources.

\subsection{Semi-quantum protocol for quantum private comparison}

Now, we would try to design a semi-quantum protocol for QPC. In which,
TP would be considered as the only quantum party (i.e., possess quantum
resources) while both Alice and Bob are classical parties. In convention,
a classical party can only access and process classical piece of information,
i.e., can only work in the computational basis \cite{talmor2007,talmor2007-1,talmor2,Med-SQKD}.
Thus, being classical user, Alice and Bob can only measure and prepare
qubits in computational basis. Additionally, they can reflect the
qubits without disturbing them. The first semi-quantum protocol of
key distribution (SQKD) was introduced in 2007 \cite{talmor2007,talmor2007-1}.
In which an unconditionally secure quantum key was shared between
a classical and quantum party by using measure and resend \cite{talmor2007,talmor2007-1,talmor2}
and permutation of particles \cite{talmor2} techniques to avoid eavesdropping.
In what follows, we refer to these two equivalent techniques (i.e.,
the schemes discussed in Refs. \cite{talmor2007,talmor2007-1,talmor2})
for eavesdropping checking as semi-quantum subroutine. The protocol
proposed in Ref. \cite{talmor2007-1} was widely discussed and was
followed by a set of SQKD protocols (\cite{Med-SQKD} and references
therein), semi-quantum DSQC \cite{sdqsc}, etc. A protocol for semi-quantum
private comparison (SQPC) is described below. In the protocol proposed
below, we will assume that Alice and Bob share a secure key $K_{AB}$,
which has been distributed/produced using SQKD protocol proposed in
Ref. \cite{Med-SQKD}, in which Alice and Bob restricted to classical
resources can share a quantum key \cite{Med-SQKD}. We further assume that Alice (Bob)  prepares an unconditionally secure key $K_{AT}$   ($K_{BT}$) in collaboration with quantum enabled TP using a semi-quantum key agreement (SQKA) protocol  \cite{SQKA}. In other words,
we assume that Alice (Bob) knows $K_{AB}$ and $K_{AT}$  ($K_{AB}$ and $K_{BT}$) before the start of the following
protocol for semi-quantum private comparison. 
\begin{description}
\item [{SQ1:}] Same as OSB1, with the only difference that the number of
EPR states prepared is $\approx8N.$ 
\item [{SQ2:}] TP sends the two sequences $S_{A}$ and
$S_{B}$\textcolor{black}{{} to} the classical users Alice and Bob,
respectively. As mentioned beforehand, the classical users can only
perform two operations, i.e., either ``measure and resend''\textbf{
}in the computational basis or\textbf{ }``reflect'' the qubit without
any change.\textbf{ } 
\item [{SQ3:}] Both the users measure half of the received qubits (i.e.,
$\approx4N$ qubits) randomly in computational basis and replace them
with the freshly prepared qubits in the same basis in accordance with
the measurement outcome. They also store the bit values obtained in
their measurement outcome corresponding to each measured qubit. 
\item [{SQ4:}] After receiving all the qubits from Alice and Bob, TP performs
Bell measurements on the received pairs of qubits (i.e., $S_{A}^{i}$
and $S_{B}^{i}$). A particular measurement would yield one of the
Bell states. If TP's measurement yields the same Bell state as was
initially prepared by him, he announces 0, otherwise he announces
1. \\

\begin{table}
\centering{}%
\begin{tabular}{|c|c|c|c|l|}
\hline 
 & Alice's  & Bob's  & \multicolumn{1}{c|}{TP's measurement} & outcome \tabularnewline
\hline 
 & choice  & choice  & Initial state $|\psi^{+}\rangle$  & Initial state $|\phi^{+}\rangle$\tabularnewline
\hline 
Case 1  & R  & R  & $|\psi^{+}\rangle$  & $|\phi^{+}\rangle$\tabularnewline
\hline 
Case 2  & M  & R  & $|\psi^{\pm}\rangle$  & $|\phi^{\pm}\rangle$\tabularnewline
\hline 
Case 3  & R  & M  & $|\psi^{\pm}\rangle$  & $|\phi^{\pm}\rangle$\tabularnewline
\hline 
Case 4  & M  & M  & $|\psi^{\pm}\rangle$  & $|\phi^{\pm}\rangle$\tabularnewline
\hline 
\end{tabular}\protect\protect\protect\caption{\label{tab:SQ-PC} All possible cases that may appear in a semi-quantum
private comparison scheme are summarized in this table. Here, both
the classical users can either choose to measure or reflect half of
the qubits, randomly and independently. We have explicitly shown all
the cases when the initial state was $|\psi^{+}\rangle$ or $|\phi^{+}\rangle$.
It is only Case 4, which is of interest, because only this case leads
to successful key sharing between Alice and Bob. Case 1 is used for
eavesdropping checking, i.e., TP's announcement ``1'' for the qubits
corresponding to Case 1 may be considered as the signature of disturbance
caused by Eve.}
\end{table}

Out of $\approx8N$ initial Bell states, in SQ3, both Alice and Bob
randomly measure one half of the received qubits. Therefore, for a
particular pair of qubits ($i$th Bell state, which was prepared initially),
there are four possible cases as listed below and summarized in Table
\ref{tab:SQ-PC}. \\
 \textbf{Case 1:} Both Alice and Bob reflect the incoming qubits undisturbed.
This case may not be useful in sharing a key between Alice and Bob,
but it can be used for eavesdropping checking. As in the absence (presense)
of an eavesdropper and/or noise the initial state prepared by TP and
the final state obtained by him as the outcome of his Bell measurement
should be the same (different in 75\% cases). \\
 \textbf{Case 2 and Case 3:} Either Alice or Bob measures the qubit
while the other party reflects it. All these cases will be discarded
as these cases are neither useful in sharing key nor for eavesdropping
checking.\\
 \textbf{Case 4:} Both Alice and Bob perform measurements in computational
basis on the qubits received by them, and store the measurement outcome
(0 or 1) to be used in the future. This case would lead to a key.

\item [{SQ5:}] After the announcement of TP, both Alice and Bob disclose
the coordinates of the qubits they have measured. Subsequently, both
of them make two separate lists corresponding to qubits falling in
Cases 1 and 4. Then, Alice and Bob check eavesdropping using semi-quantum
subroutine using qubits of Case 1. In which, both the classical parties
ensure that TP's measurement outcome is 0 when both have reflected
the qubits (Case 1). In this case, about one-fourth of the initial
states (i.e., $\approx2N$ Bell states) will be used for eavesdropping
checking. If the errors are found to be lower than a tolerable limit,
they proceed to the next step, otherwise they discard the scheme,
and restart from SQ1. \\
 On the successful completion of this step (i.e., if the error rate
is found to be less than the tolerable limit), Alice and Bob obtain
$K_{A_{2N}}$ and $K_{B_{2N}}$, respectively, where $K_{A_{2N}}$
and $K_{B_{2N}}$ are correlated strings of length $\approx2N$-bits. 
\item [{SQ6:}] On request of Alice and Bob, TP discloses his initial choices
of Bell states for $N$ cases, that are selected randomly. Subsequently,
using the values of $K_{A_{2N}}^{i}$ and $K_{B_{2N}}^{i}$, Alice
and Bob compare the parity values for the Bell states prepared by
TP and the bit values obtained by them for all $N$ cases. If it matches,
they proceed, otherwise they abort. \\
 This would check whether TP had genuinely prepared a Bell state or
not, which in turn would ensure the correlation in the reduced $N$-bit
strings $K_{A}$ and $K_{B}$. 
\item [{SQ7:}] Same as OSB6 with the only difference 
that Alice (Bob) encrypts her (his) secret $M_{A}$ ($M_{B}$) with her (his) key string
$K_{A}$ ($K_{B}$), key shared with Bob (Alice) $K_{AB}$, and TP $K_{AT}$ ($K_{BT}$) by using an exclusive-OR operation
to obtain $C_{A}^{i}=\left(M_{A}^{i}\oplus K_{A}^{i}\oplus K_{AB}^{i}\oplus K_{AT}^{i}\right)$
($C_{B}^{i}=\left(M_{B}^{i}\oplus K_{B}^{i}\oplus K_{AB}^{i}\oplus K_{BT}^{i}\right)$.
before announcing the strings $C_{A}$ and $C_{B}$
separately to TP via a public channel. 
\item [{SQ8:}] Same as OSB7. 
\item [{SQ9:}] Same as OSB8 with a minor change in TP's calculation of $R^{i}$, which is exclusive-OR result of $C_{A}$,
$C_{B}$, $C_{TP}$, $K_{AT}$, and $K_{BT}$ as $R=\left(C_{A}\oplus C_{B}\oplus C_{TP}\oplus K_{AT}\oplus K_{BT}\right)$.\color{black}
\end{description}

\section{Analysis of security and efficiency of the proposed protocols\label{sec:Analysis-of-security-efficiency}}

Before we analyze the security of the proposed protocols, it would
be beneficial to summarize the information known (or unknown) to each
party. To be specific, in both the protocols proposed in the last
section, TP has the information of the initial Bell states, which
is unknown to both Alice and Bob. This ensures 1 bit of ignorance
for both Alice and Bob. This is so because the initial Bell states
can be distinguished as parity 0 ($|\psi^{\pm}\rangle$) state and
parity 1 ($|\phi^{\pm}\rangle$) state. For a parity $0$ (1) state
the final key bits obtained by Alice and Bob would be the same (complementary
of each other). Hence, the ignorance due to the lack of knowledge
about the initial Bell state is 1 bit. Interestingly, the initial
choice of Bell states also ensure 1 bit of ignorance for TP as he
has knowledge about the parity of the initial Bell state, but he does
not know whether Alice's (or Bob's) measurement outcome was 0 or 1.
Further, Alice (Bob) has the shared quantum key $K_{AB}$ and correlated
key $K_{A}$ ($K_{B}$), which are not known to TP. It is evident
that anyone interested to know the secrets of Alice and/or Bob ($M_{A}$
and/or $M_{B}$) would require all three parts of information.

Firstly, it should be ensured that a sequence of qubits transmitted
via a quantum channel has been received by the receiver in a secure
manner as an unintended external attacker may employ an intercept-resend
or entangle-measure attack. To ensure that any such eavesdropping
attack is detected, one needs to check half of the travel qubits.
This is why, $N$ decoy qubits are inserted in each sequence sent
to Alice and Bob by TP in OSB1 in the orthogonal-state-based protocol
proposed here. This is essential because it would provide an upper
bound on the errors in the qubits not checked for eavesdropping, and
the error would become asymptotically small for large values of $N$
\cite{nielsen}. In the semi-quantum protocol, the security against
an outsider attacker has been obtained from the qubits reflected by
both Alice and Bob (already discussed as Case 1 in SQ4). Specifically,
in this particular case, in the absence of an eavesdropper and/or
noise the initial state prepared by TP and the final state obtained
by him as the outcome of his Bell measurement should be the same.
However, the presence of noise and eavesdropper can lead to other
Bell states. Thus, if TP prepares $|\psi^{+}\rangle\,(|\phi^{+}\rangle)$
and obtains any other state at the end, he can conclude that there
is an eavesdropper or noise, but the converse is not true. However,
since $N\gg1,$ any effort of eavesdropping will always be detected.
Further, in SQ6, Alice and Bob check the parity of their bits in half
of the cases with the TP's announcement of the initial choice of Bell
state. This would check whether TP had genuinely prepared a Bell state
or not, which in turn would ensure the correlation in the reduced
$N$-bit strings $K_{A}$ and $K_{B}$. This disclosure will also
be relevant in some of the intercept and resend attacks of dishonest
Alice or Bob.  Instead of trying to extract the useful information, Eve may also choose to disturb the protocol. For example, in the orthogonal-state-based QPC protocol, she may apply a Pauli operation $iY$ on all the qubits traveling from TP to Alice without disturbing Bob's qubits. Alice would not be able to detect this attack using the GV subroutine as in OSB2  all the decoy Bell states were prepared as $|\psi^{+}\rangle$ and those would remain unchanged after this operation, while parity of the qubits used for preparing the key will change completely (when compared with that of the Bell state prepared by TP). Specifically, as a consequence of this attack, TP would be forced to announce incorrect results in two cases: (i) when every bit in $M_{A}$ and $M_{B}$ are the same, and (ii) when every bit in $M_{A}$ and $M_{B}$ are different, and  Alice and Bob would not succeed  to detect this attack. As mentioned previously, Eve cannot extract any information encoded by Alice or Bob through this attack, but she may force Alice and Bob to reach at a wrong conclusion, which is not desirable. This undesirable situation can be circumvented by introducing an additional security checking (as mentioned in OSB4),  where Alice and Bob check correlations in a part of the secure strings shared by them with the help of TP.  This correlation check would reveal whether Eve has adopted the above mentioned attack strategy. \color{black}

Now, we can consider various possible attacks by each party to gain
the information not accessible to them. For example, TP can try to
take benefit of state preparation by preparing a state other than
Bell state to gain advantage by entanglement swapping attack \cite{zhang-qinp-2013-cryptanalysis-yang-2013}.
In fact, he can prepare an arbitrary quantum state to exploit the
state preparation, and try to extract correlated keys $K_{A}$ and
$K_{B}$. However, the unconditional security of the quantum key $K_{AB}$
would maintain security against any such attempt as both Alice and
Bob encrypt their messages regarding asset information $M_{A}$ and
$M_{B}$, respectively using the quantum key $K_{AB}$. In other words,
it can be viewed as Alice and Bob performing communication in a secure
manner by using a quantum key and the security of message can be attributed
completely to the security of the quantum key. Additionally, Alice
(Bob) may wish to extract Bob's (Alice's) key $K_{B}$ ($K_{A}$)
by designing an intercept and resend attack \cite{zhang-qinp-2013-cryptanalysis-yang-2013}.
The eavesdropping checking mechanism adopted in both the protocols
(by GV and semi-quantum subroutines, respectively) ensures security
against this kind of an attack. It may be noted that the eavesdropping
checking procedure adopted here would provide security against both
insider and outsider attackers.

A modified intercept and resend attack strategy by a classical Alice
(Bob) in SQPC scheme could be to intercept all the qubits sent to
Bob (Alice) and keep them in a memory. Though, possession of a quantum
memory by a classical party is beyond his/her limits, but it is an
interesting scenario to be investigated. Suppose Alice measures half
of the qubits which TP has sent to her, and follows the same for the
corresponding Bob's qubits in her possession. Depending upon the outcomes
of her measurement, Alice prepares new qubits and sends (the same
number of qubits TP would have sent) them to Bob after adding the
remaining auxiliary qubits. Bob proceeds with the protocol as expected,
then Alice intercepts all the qubits Bob returns to TP and replaces
them with the qubits (initially sent by TP to Bob) in her possession.
Using this approach, (if $K_{BT}$ is not used) she will get $K_{B}$ completely as it will be
prepared using a subset of the string of qubits, she had measured
and sent to Bob.  Using a modification of this attack, she may decide to not to intercept qubits from TP to Bob and still perform a similar attack. Specifically, Alice (or equivalently Bob) can also extract all the information of Bob by intercepting all the qubits that Bob has sent to TP and measuring the pair qubits of each qubit that Alice had measured. None of these two attacks can be detected using the semi-quantum subroutine alone. However, these attacks can be circumvented by allowing Alice and Bob to prepare individual  quantum keys with quantum enabled TP using SQKA protocol (as proposed in Ref. \cite{SQKA}). The use of SQKA scheme instead of a SQKD scheme would allow the protocol to succeed even when the TP is not honest. Now, using SQKA scheme, if Alice and TP (Bob and TP) prepare a key $K_{AT}$ ($K_{BT}$), and Alice and Bob use  these additional keys $K_{AT}$ and $K_{BT}$ to encode their messages $M_{A}$ and $M_{B}$, respectively (as described in SQ7), then the above mentioned attacks can be circumvented. This is easy to observe that in the modified scheme, by performing one of the above mentioned attacks though Alice would obtain $K_{B}$,  which would not help her to deduce $M_{B}$ as she is completely ignorant about $K_{BT}$.  \color{black}

Additionally, it is worth noting that the SQPC protocol intrinsically
uses a scheme for QKD to obtain its security against the untrusted
TP and external attackers. Thus, a QKD scheme, which is unconditionally
secure \cite{vaidman-goldenberg,GV-experiment,N09,Med-SQKD} and composable
\cite{composibility}, is used here to provide security from attackers
other than Alice and Bob (primarily from the malicious TP). To ensure
this security, Alice and Bob are required to be honest as they do
not have a better strategy (to protect their individual secret from
the malicious TP) than to honestly share a key. Here, we may note
that composability \cite{composability2} of QKD plays a crucial role
here as it specifies additional security criteria that must be fulfilled
in order for QKD to be composed with other tasks to form a larger
application like QPC. Moreover, we may note that the criteria for
composability would be more stringent in a situation (like QPC) that
involves mutually mistrustful parties (see \cite{composability3}
for details). Further, unlike the orthogonal-state-based protocol
for QPC which uses QKA involving both dishonest parties, in SQPC,
Alice and Bob share a quantum key using a semi-quantum KD protocol
\cite{Med-SQKD}, which does not allow any one of them to solely decide
the entire key (thus, the QKD scheme used here can be viewed as a
weaker version of QKA) as the QKD scheme considered here puts both
Alice and Bob on the same footing. This is so because the untrusted
TP prepares Bell states and shares among Alice and Bob to form a symmetric
key. Consequently, neither Alice nor Bob can control the whole key.
Finally, iff both of them encrypt their information with this symmetric
key $K_{AB}$, the result at the TP's end will have a $R$ which would
be free from the contribution of key. This fact has been exploited
in some of the recent proposals where the competing parties share
a quantum key \cite{chi-type-state-ijtp-2012-liu.,zhang-qinp-2013-cryptanalysis-yang-2013,sun-2015-qinp}
or a random number \cite{yang-2013-qinp-comment-on-Hwang-2012} honestly
for security against untrusted TP.

The qubit efficiency of the proposed QPC protocols can be calculated
using the quantitative measure proposed in Refs. \cite{defn-of-qubit-efficiency}.
The efficiency $\eta=\frac{c}{q+b},$ where $c$ number of classical
bits are transmitted using the total number of $q$ qubits, and $b$-bits
classical communication is involved. It should be noted that only
the classical communication used for decoding the message is considered,
not the classical communication required for eavesdropping checking.
In both the orthogonal-state-based QPC and SQPC protocols, Alice and
Bob share their $N$-bit secrets, contributing equally in $c=2N$.
The orthogonal-state-based QPC involves $4N$ qubits, which are shared
in a secure manner using $4N$ number of additional decoy qubits.
Finally, Alice and Bob announce $C_{A/B}$ of $N$-bits each which
is followed by 1 bit of classical communication by TP (i.e., the announcement
that discloses whether the assets are equal (say, 0) or not (say,
1)). Additionally, Alice and Bob also share a quantum key and the
qubits and bits used to obtain this shared key should also be counted
in the computation of efficiency. Here, we have chosen orthogonal-state-based
QKA scheme \cite{QKA-cs} for considering the resources involved in
sharing the quantum key. Using Shukla et al.'s scheme \cite{QKA-cs},
and exploiting the dense coding capacity of the quantum channel used,
an $N$-bit quantum key can be shared by two parties using $4N$ qubits
($2N$ as quantum channel and $2N$ decoy qubits), which involves
$3N$-bit classical communication. Therefore, total number of qubits
used are $q=12N$\textbf{ }with additional classical communication
\textbf{$b=5N+1$}. Thus, the efficiency of the orthogonal-state-based
QPC protocol would be $\eta=\frac{2N}{17N+1}$, which becomes 11.76\%
for large $N$.

Similarly, in case of SQPC protocol, $16N$ qubits are initially prepared
by TP, subsequently $8N$ qubits are prepared by Alice and Bob ($4N$
by each) as a replacement of the qubits measured by them. TP announces
$8N$ bit measurement outcomes which Alice and Bob use to generate
their keys with the help of $4N$ bit of classical messages available
with each of them. Finally, Alice and Bob both share an $N$-bit encrypted
messages which helps TP to obtain and announce 1 bit final result.
Additionally, both classical parties also share a secure quantum key
\cite{Med-SQKD} for which they use $24N$ bits (i.e., TP initially
prepares $16N$ qubits, and Alice (Bob) prepares additional $4N$
($4N$) qubits). Similar to the SQPC protocol, during key distibution TP also announces $8N$-bit
information which is followed by $4N$-bit announcement
by both Alice and Bob. It is important to note that Alice and Bob
keep the bit values only for specific announcements by TP. Here, we
have considered the asymptotic case where equal probabilities for
favorable and unfavorable measurement outcomes of TP has been considered. In the present SQPC protocol, TP had also shared one unconditionally secure quantum key with each classical user using a SQKA scheme \cite{SQKA}, which involves $5N$ qubits and $5N$ bits for each $N$-bit shared key. 
Using all these values, the efficiency of the SQPC protocol is $\eta=\frac{2N}{102N+1}$.
For large values of $N$, $\eta\approx1.96\%$. Thus, the protocol
1 proposed here is more efficient than the protocol 2, which is not
surprising as we have considered two classical users in the second
protocol.

Here, it would be worth mentioning that in analogy with the classical
communication involved in eavesdropping checking process (which is
not counted in computation of $\eta)$, we have not counted the classical
bits that are used to check correlations between $K_{A}$ and $K_{B}$.

\section{Effect of different noise models on the proposed protocols\label{sec:Effect-of-different-noise}}

In this section, we aim to study the effect of a set of noise models
on the feasibility of both the QPC schemes proposed here. For this
we will use the operator sum representation of a transformed quantum
state ($\rho^{\prime}$), initially prepared in $\rho=|\psi\rangle\langle\psi|$,
which is 
\begin{equation}
\rho^{\prime}=\stackrel[i]{n}{\sum}K_{i}\rho K_{i}^{\dagger}.\label{eq:noise-affected}
\end{equation}
Here, the operators satisfying $\stackrel[i]{n}{\sum}K_{i}^{\dagger}K_{i}=\mathbb{I}$
ensure the trace-preserving nature of the quantum channel. In what
follows, first we will describe the Kraus operators of the noise channels
and mathematical formulation used in the present letter. Subsequently,
we will analyze both of the proposed schemes under the effect of various
noise models. Specifically, we will consider the amplitude damping
(AD) channel, bit flip channels, phase flip channels and depolarizing
channels. Let us first describe an AD channel, which causes the loss
of energy from the system to its surrounding considered as a reservoir
at vacuum. 
\begin{description}
\item [{1.}] \textbf{Amplitude damping: }The Kraus operators for AD channels
\cite{nielsen} are $E_{0}^{AD}=\left(\begin{array}{cc}
1 & 0\\
0 & \sqrt{1-p}
\end{array}\right)$ and $E_{1}^{AD}=\left(\begin{array}{cc}
0 & \sqrt{p}\\
0 & 0
\end{array}\right)$. Here, and in what follows, $p$ is the probability of error due
to the specific noise discussed. 
\end{description}
We will also study the effect of Pauli noise, which affects the qubit
independent of its initial state with a certain probability and leaves
it unaffected with the remaining probability. The error caused due
to this kind of noise can be studied as bit flip (BF), phase flip
(PF), and depolarizing channel (DC) noise. 
\begin{description}
\item [{2.}] \textbf{Bit flip channels: }The BF channel, which flips the
qubit with probability $p$ and leaves it unchanged with remaining
probability, can be given by the Kraus operators \cite{nielsen} $E_{0}^{BF}=\sqrt{1-p}\left(\begin{array}{cc}
1 & 0\\
0 & 1
\end{array}\right)$ and $E_{1}^{BF}=\sqrt{p}\left(\begin{array}{cc}
0 & 1\\
1 & 0
\end{array}\right)$. 
\item [{3.}] \textbf{Phase Flip channels: }Similar to the BF channels,
the Kraus operators for a phase flip channel are given by \cite{nielsen}
$E_{0}^{PF}=\sqrt{1-p}\left(\begin{array}{cc}
1 & 0\\
0 & 1
\end{array}\right)$ and $E_{1}^{PF}=\sqrt{p}\left(\begin{array}{cc}
1 & 0\\
0 & -1
\end{array}\right)$, which flips the phase of the qubits with $p$ probability while
identity acts with the remaining probability. It is interesting that
the effect of phase damping noise on certain scheme can be reduced
from the obtained results for phase flip channels (\cite{vishal}
and references therein), so we have not explicitly studied the phase
damping kind of noise. 
\item [{4.}] \textbf{Depolarizing channels: }If the noisy channels leaves
the state unchanged with a certain probability and completely mixed
with the remaining probability, the Kraus operators for such a channel
would be \cite{nielsen} $E_{0}^{DC}=\sqrt{1-p}\left(\begin{array}{cc}
1 & 0\\
0 & 1
\end{array}\right)$, $E_{1}^{DC}=\sqrt{\frac{p}{3}}\left(\begin{array}{cc}
0 & 1\\
1 & 0
\end{array}\right)$, $E_{2}^{DC}=\sqrt{\frac{p}{3}}\left(\begin{array}{cc}
0 & -i\\
i & 0
\end{array}\right)$, and $E_{3}^{DC}=\sqrt{\frac{p}{3}}\left(\begin{array}{cc}
1 & 0\\
0 & -1
\end{array}\right).$ 
\end{description}
In both the QPC schemes proposed here, TP initially prepares one of
the Bell states and shares that with two distant parties. Therefore,
$\rho$ will be a two qubit density matrix evolving under two independent
quantum channels. Mathematically, the transformed density matrix can
be written as 
\begin{equation}
\rho^{\prime}=\sum_{i,j}E_{i}^{k}\left(p_{1}\right)\otimes E_{j}^{l}\left(p_{2}\right)\rho\left(E_{i}^{k}\left(p_{1}\right)\otimes E_{j}^{l}\left(p_{2}\right)\right)^{\dagger},\label{eq:noise-scheme-1}
\end{equation}
where $p_{i}$ corresponds to the probability of decoherence in the
$i$th qubit and $E_{i}^{k}$ are the Kraus operators of the noisy
channel acting on the particular qubit, with $k,l\in\left\{ AD,BF,PF,DC\right\} $.
To quantify the effect of noise we will use a distance based measure,
fidelity, between the quantum state received by the receivers (Alice
and Bob here) after the effect of a specific noise channel and the
quantum state expected in the ideal condition, when noisy channel
do not alter the quantum state. In fact, in the ideal condition, the
state at the receivers end should be the same prepared by TP. Therefore,
the fidelity between the initial ($\rho$) and final ($\rho^{\prime}$)
state is 
\begin{equation}
F=\langle\psi|\rho^{\prime}|\psi\rangle.\label{eq:fidelity}
\end{equation}
Here, it is also worth mentioning that the fidelity expression used
here is square of the conventional fidelity expression but is also
used widely as a measure (\cite{RSP-with-noise,crypt-switch,fdly}
and references therein). Now, we will discuss the effect of all these
noise models on the proposed schemes.

\subsection{Effect of noise on the protocols}

It is noteworthy that the qubits used by Alice and Bob to prepare
their respective keys undergo the same fate in both the schemes. Therefore,
the discussion what follows is applicable to both orthogonal-state-based
QPC and SQPC protocols. Specifically, when both the qubits of the
initial state $|\psi^{\pm}\rangle$ are subjected to AD noise the
compact analytic expression of the obtained fidelity is

\begin{equation}
F=\frac{1}{4}\left(2+2\sqrt{\left(1-p_{1}\right)\left(1-p_{2}\right)}-\left(p_{2}+p_{1}\right)+2p_{1}p_{2}\right),\label{eq:AD-AD}
\end{equation}
where $p_{1}$ and $p_{2}$ are the probabilities of errors in the
first and second qubits, respectively. The same calculated for the
different choice of initial state, i.e., $|\phi^{\pm}\rangle$is 
\begin{equation}
F=\frac{1}{4}\left(\sqrt{1-p_{1}}+\sqrt{1-p_{2}}\right){}^{2}.\label{eq:AD-AD PHIPLUS}
\end{equation}
In what follows, the same fidelity expressions are obtained irrespective
of the initial choice of Bell state under different noisy environment.

Similar to the earlier case, when the first qubit evolves under the
effect of AD channel, whereas the second qubit is affected by BF channel
the calculated fidelity turns out to be

\begin{equation}
F=\frac{1}{4}\left(-2\left(1+\sqrt{1-p_{1}}\right)\left(-1+p_{2}\right)+p_{1}\left(-1+2p_{2}\right)\right).\label{eq:AD-BF}
\end{equation}
Similarly, when the first qubit is subjected to AD noise and the second
qubit evolves under the effect of PF and DC noise models, we obtain
the fidelity expressions as

\begin{equation}
F=\frac{1}{4}\left(2+2\sqrt{1-p_{1}}-p_{1}-4\sqrt{1-p_{1}}p_{2}\right)\label{eq:AD-PF}
\end{equation}
and

\begin{equation}
F=\frac{-2p_{1}\left(-1+p_{2}\right)+\left(1+\sqrt{1-p_{1}}\right)\left(-4+3p_{2}\right)}{4\left(-2+p_{2}\right)},\label{eq:AD-DN}
\end{equation}
respectively.

Before we consider the effect of other kind of noise models (other
than AD) on the first qubit, it is worth noting that the scheme is
symmetric with respect to Alice and Bob (due to the symmetry of the
Bell states used), and the similar fidelity expressions are expected
whether the first (second) qubit is affected by AD (BF) or the first
(second) qubit is affected by BF (AD) noise. In fact, if we interchange
the values of $p_{1}$ and $p_{2}$ we can obtain the fidelity expression
of one from the other. Due to this fact, we are enlisting here only
the expressions for fidelity, for those cases which lack this kind
of symmetry (i.e., which cannot be obtained from another expression
by simply using a symmetry argument).

If the first qubit is subjected to BF noise, while the second qubit
evolves under BF, PF and DC noises, then a systematic computation
would yield the fidelity expressions as

\begin{equation}
F=1-p_{2}+p_{1}\left(-1+2p_{2}\right),\label{eq:BF-BF}
\end{equation}

\begin{equation}
F=\left(-1+p_{1}\right)\left(-1+p_{2}\right),\label{eq:BF-PF}
\end{equation}
and

\begin{equation}
F=\frac{3}{2}-2p_{1}+\frac{1-2p_{1}}{-2+p_{2}},\label{eq:BF-DN}
\end{equation}
respectively. It is interesting to report that the fidelity obtained
while both the qubits evolve under PF environment is exactly the same
as that under BF noise (\ref{eq:BF-BF}).

If we consider PF noise on the Alice's qubit and DC noise on the Bob's
qubit we obtain a closed form analytic expression of fidelity as

\begin{equation}
F=\frac{\left(-1+p_{1}\right)\left(-4+3p_{2}\right)}{2\left(2-p_{2}\right)}.\label{eq:PF-DN}
\end{equation}
Finally, when both the qubits are subjected to DC noise, the obtained
fidelity between the quantum state affected by noise and the initial
state is

\begin{equation}
F=\frac{8-6p_{2}+p_{1}\left(-6+5p_{2}\right)}{2\left(-2+p_{1}\right)\left(-2+p_{2}\right)}.\label{eq:DN-DN}
\end{equation}

\begin{figure}
\centering{}\includegraphics[scale=0.55]{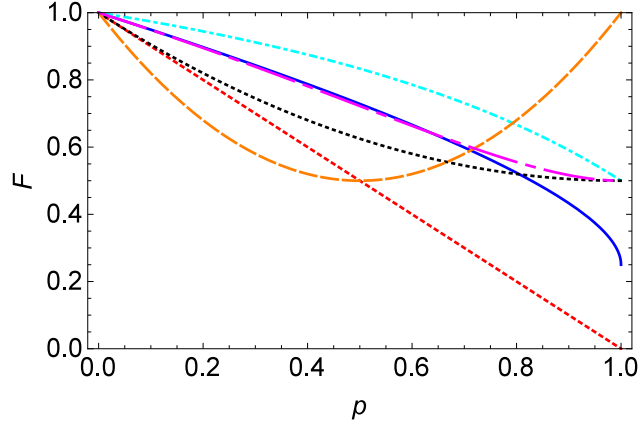}\protect\protect\protect\caption{\label{fig:qubit1or2}The fidelity variation of the quantum states
for the QPC protocols in specific cases of noisy environment is shown
here. The smooth (blue), small-dashed (red) and dot dashed (cyan)
curves correspond to the fidelity of the obtained states when only
Alice's (or Bob's) qubit is subjected to AD, BF and DC noises, respectively.
Similarly, the dotted (black), large-dashed (orange) and large-dot
dashed (magenta) lines show variation in the fidelity of the obtained
states with probability of errors, when both Alice's and Bob's qubits
are subjected to the same kind of noise with equal probabilities of
errors, i.e., AD, BF and DC noises, respectively. In both the cases
of PF channels, the same fidelity as that under BF noise is obtained.
As discussed in the text the choice of initial Bell states only matter
when both the qubits are subjected to AD noise. Here, the dotted (black)
line shows the fidelity variation for the initial state $|\psi^{\pm}\rangle$
while the small-dashed (red) line corresponds to the initial state
$|\phi^{\pm}\rangle$.}
\end{figure}

Before we proceed with the analysis of the various fidelity expressions
obtained under different combinations of noisy channels, it is worth
establishing the motivation of this study. Specifically, we have already
mentioned the fidelity between the initial and final state in the
ideal conditions is expected to be unity. For the sake of argument,
consider that one of the channel (either TP to Alice or TP to Bob)
is noiseless, then this unit fidelity falls considerably (cf. Fig.
\ref{fig:qubit1or2}). Specially, for the higher values of decoherence
rate, it even becomes null for BF or PF channels. More realistic scenario
would be where both the qubits undergo decoherence. For simplicity
of discussion, if we consider that the rate of decoherence in both
the channels is the same, then the fidelity of the obtained state
evolving under BF or PF noise turned upside down to become unity for
higher probability of errors. Though, a similar benefit appears when
the qubits evolve under AD noise. However, no such advantage is seen
in presence of the DC noise.

\begin{figure}
\centering{}\includegraphics[angle=0,scale=0.45]{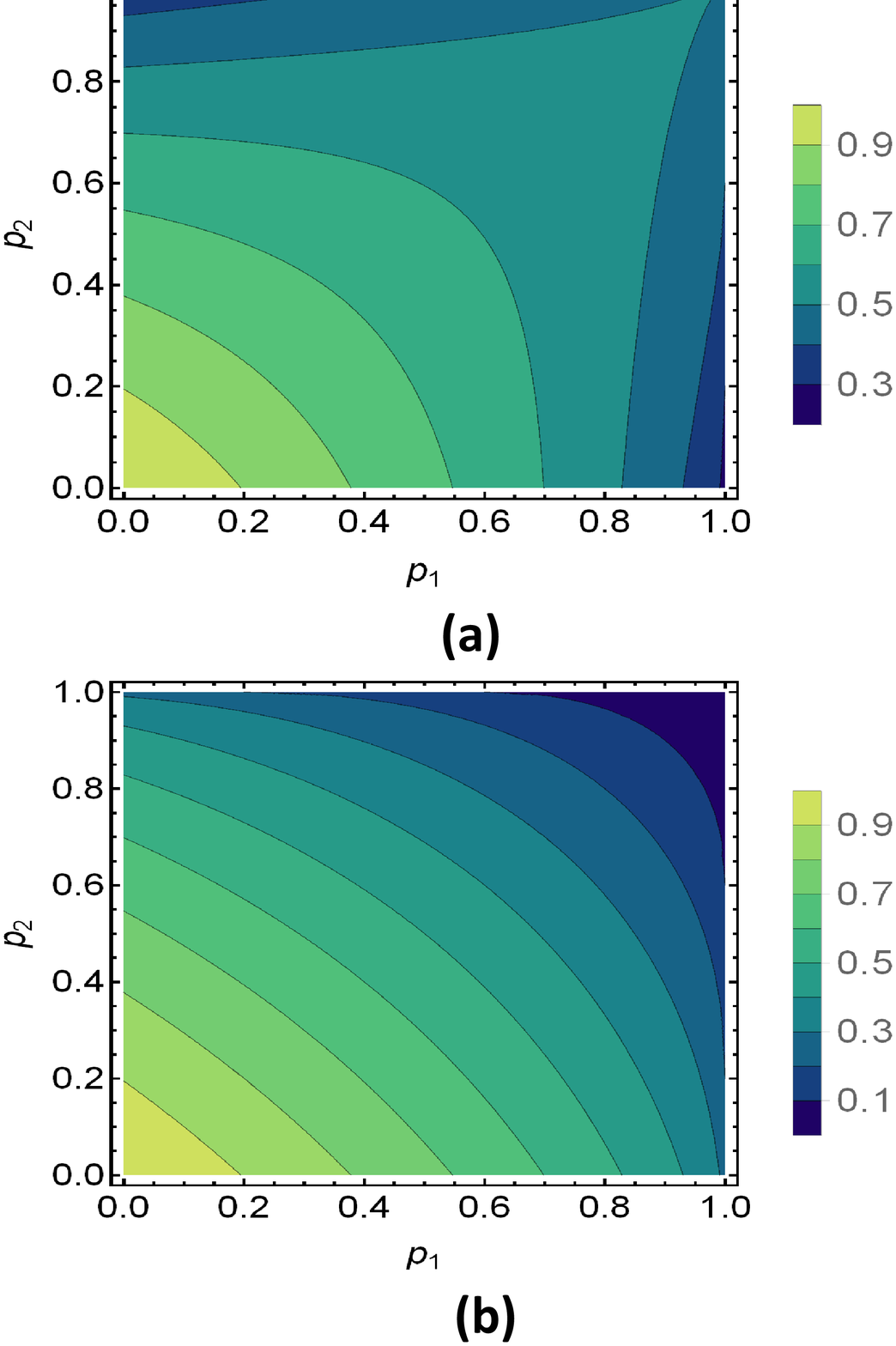} \protect\protect\protect\caption{\label{fig:AD}QPC protocols subjected to AD channels, i.e., both
the qubits evolve under AD noise. In (a) and (b), the choice of the
initial Bell state by TP is $|\psi^{\pm}\rangle$ and $|\phi^{\pm}\rangle$,
respectively.}
\end{figure}

\begin{figure}
\centering{}\includegraphics[angle=0,scale=0.50]{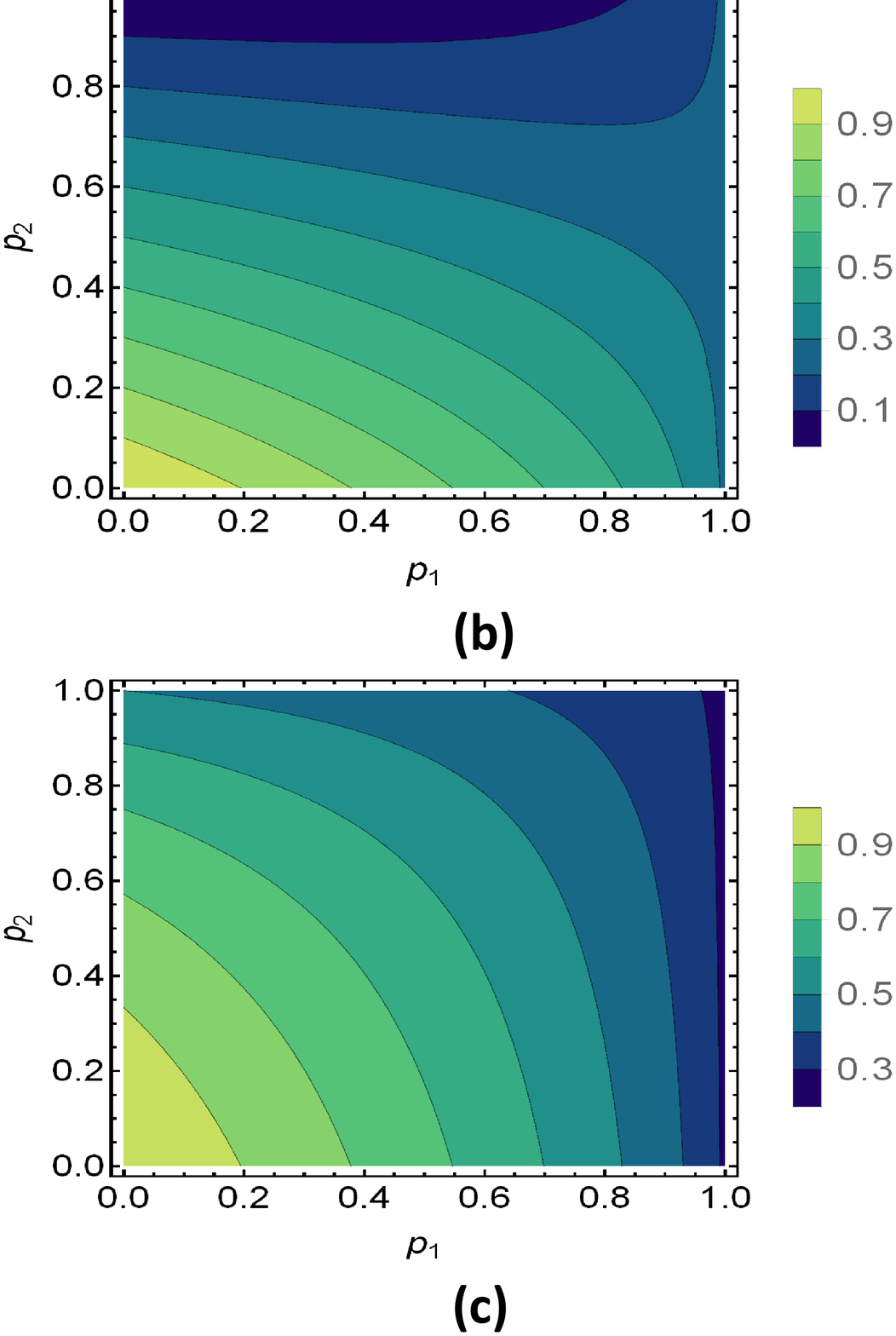} \protect\protect\protect\caption{\label{fig:AD1-Pauli2}QPC protocols subjected to a noisy environment,
when the first qubit of the Bell state (Alice's qubit) is subjected
to AD, while the second (Bob's) qubit evolves under different noisy
channels. In (a), (b), and (c), it evolves under BF, PF, and DC, respectively.}
\end{figure}

\begin{figure}
\centering{}\includegraphics[angle=0,scale=0.5]{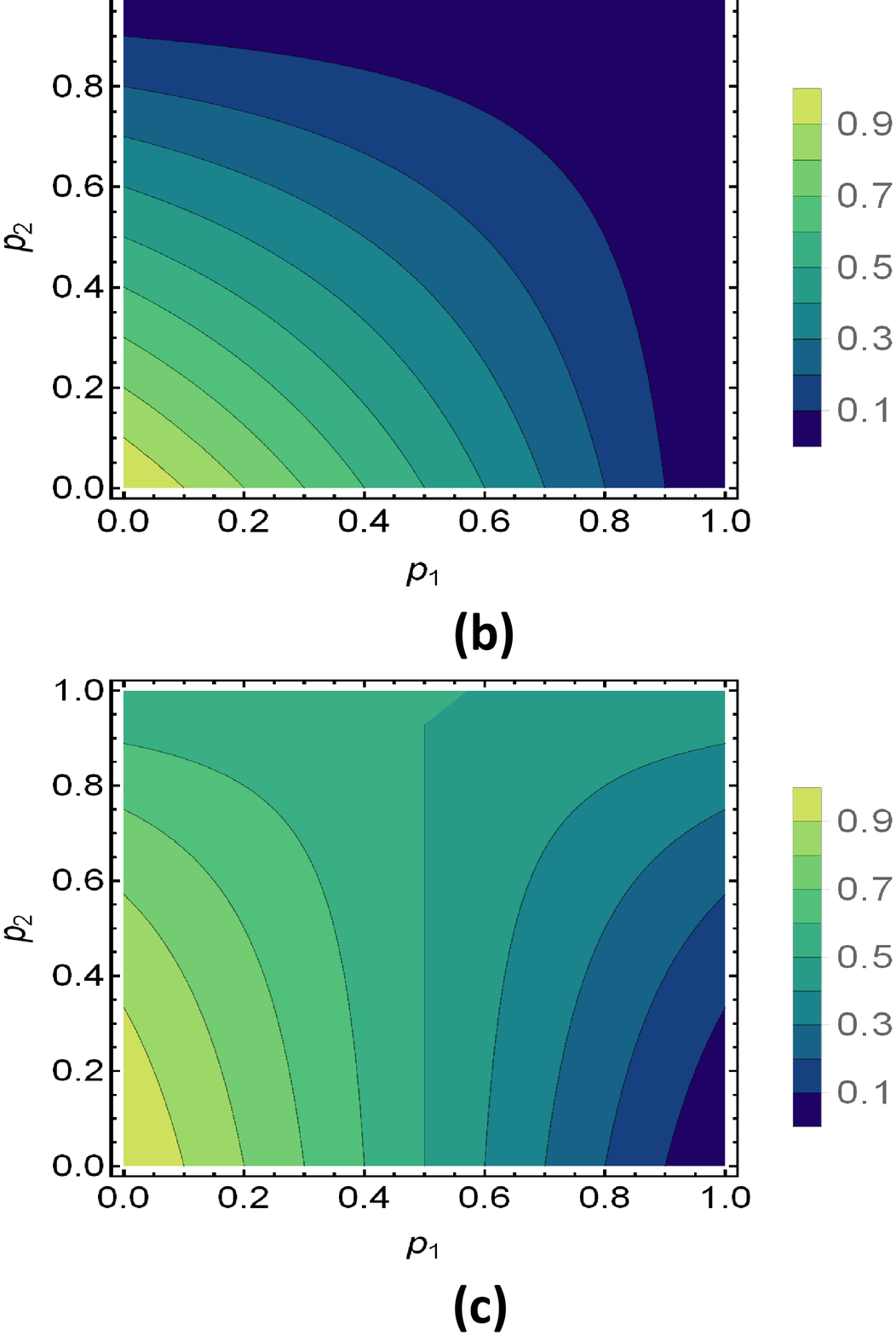} \protect\protect\protect\caption{\label{fig:BF1-Pauli2}QPC protocols subjected to noisy environment,
when the first qubit of the Bell state (Alice's qubit) is subjected
to BF, while the second (Bob's) qubit evolves under different noisy
channels. In (a), (b) and (c) Bob's qubits is affected by BF, PF,
and DC, respectively. As mentioned in the text (a) also corresponds
to the PF noise on both the qubits.}
\end{figure}

\begin{figure}
\centering{}\includegraphics[angle=0,scale=0.45]{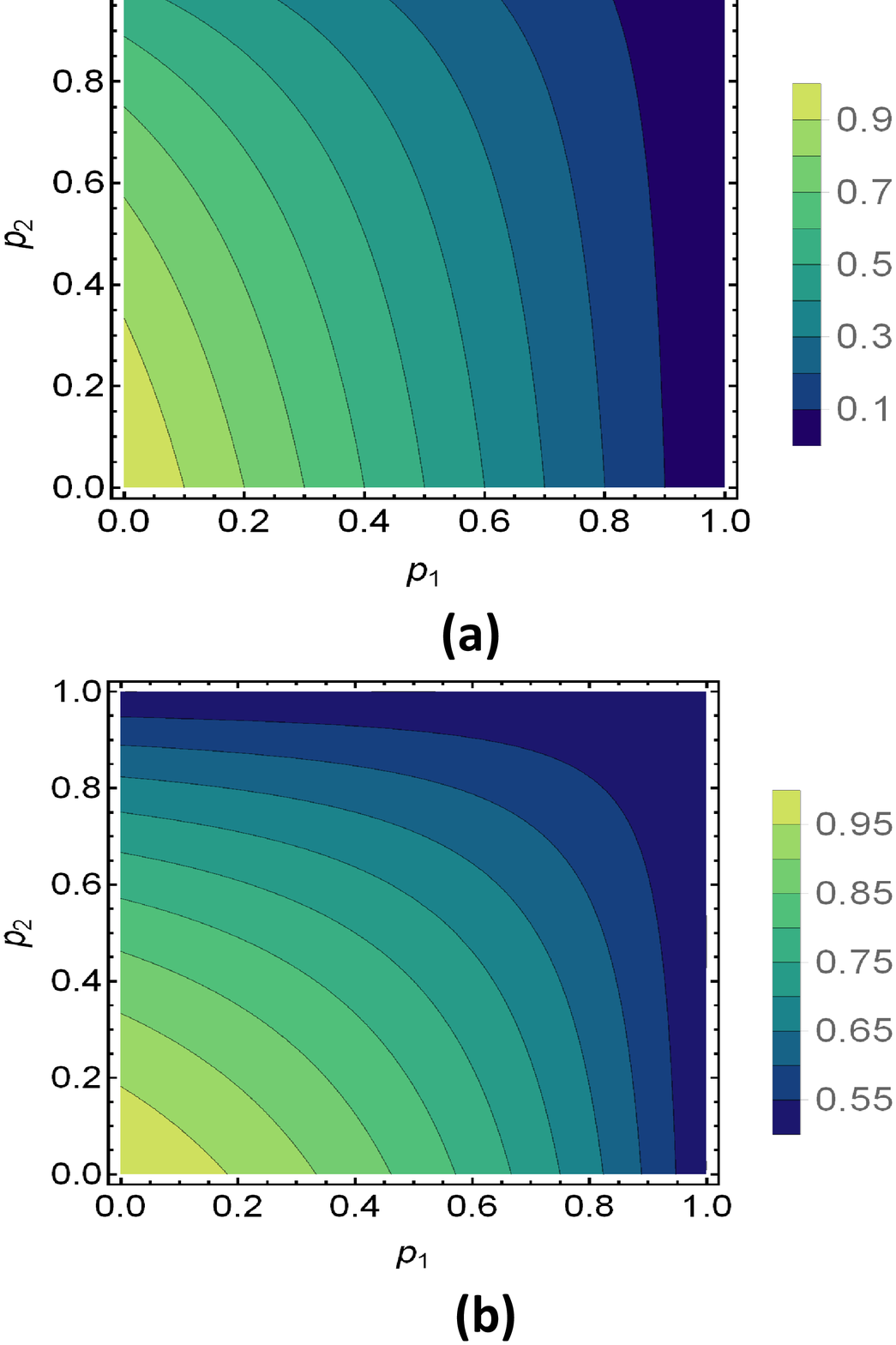}\protect\protect\protect\caption{\label{fig:PF-DC}(a) QPC protocol subjected to a noisy environment,
when the first qubit of the Bell state (Alice's qubit) is subjected
to PF while the second (Bob's) qubit evolves under DC. (b) Similarly,
when both the qubits of the Bell state (i.e., Alice's and Bob's qubits)
evolve under the effect of DC noisy environment.}
\end{figure}

We have already reported ten fidelity expressions (Eqs. (\ref{eq:AD-AD})-(\ref{eq:DN-DN}))
calculated for the quantum states shared between all the parties and
evolving under different noise channels. All the analytic expressions
depend on two independent variables (probability of error). Therefore,
variation of fidelity with these variables can be explicitly illustrated
using contour plots. Specifically, in Figs. \ref{fig:AD}-\ref{fig:PF-DC},
we have shown the contour variation of Eqs. (\ref{eq:AD-AD})-(\ref{eq:DN-DN}).
In Fig. \ref{fig:AD}, the fidelity as a function of both the probabilities
$p_{i}$s is shown when the initial choice of Bell state by TP was
$|\psi^{\pm}\rangle$ and $|\phi^{\pm}\rangle$ in (a) and (b), respectively.
A gradual decay in fidelity with each independent probability appears
in (a), and when both the decoherence rates are higher, relatively
better results in terms of fidelity are visible. Whereas (b) gives
a contrasting picture, and fidelity is found to fall continuously
with both the probabilities. The comparative analysis reveals that
$|\psi^{\pm}\rangle$ gets less affected than $|\phi^{\pm}\rangle$.
It may be easily observed in both of these plots (i.e., Figs. \ref{fig:AD}
(a) and (b)) that the contour variation is symmetric along a diagonal
from lower left to top right corner due to the fact that the same
noisy channel is affecting both the qubits. It will remain valid in
all such plots to follow (cf. Figs. \ref{fig:BF1-Pauli2} (a) and
\ref{fig:PF-DC} (b)) where the same noise acts on both qubits and
symmetric nature is visible in the contour plots.

If we consider that quantum channel for TP to Alice is characterized
as AD while Bob's qubit is affected by one of the BF, PF or DC noise,
then the symmetry observed in Fig. \ref{fig:AD} is lost (cf. Fig.
\ref{fig:AD1-Pauli2} (a), (b), and (c), respectively). The variations
of fidelity in all these cases show that it is not sufficient to characterize
one quantum channel, as the type of noise applied to the second qubit
is also relevant. If we look closely at/in these three contour plots
(i.e., Fig. \ref{fig:AD1-Pauli2} (a), (b), and (c)), we can see the
same seven contour lines at the starting point at/on the $X$-axis
undergoing different variation due to the effect of the noisy channel
of the second qubit. Among all these noise models, DC noise is found
to minimally affect the fidelity, while PF noise is found to have
most devastating effects.

Another possibility is that the first qubit is subjected to BF noise.
In Fig. \ref{fig:BF1-Pauli2} (a), another qubit is also traveling
through the BF channel. Interestingly, the degrading of fidelity with
an increase in probability of error in one of the channels can be
controlled by increasing the decoherence rate for the other channel.
The same intercepts on the $X$-axis for all three plots correspond
to the same type of noise acting on the first qubit. Unlike the second
qubit evolving under BF channel, PF noise does not check the decay
in fidelity (cf. Fig. \ref{fig:BF1-Pauli2} (b)). In fact, least fidelity
is obtained in Fig. \ref{fig:BF1-Pauli2} (b) when compared among
the second qubit evolving under BF, PF, and DC noise. While the second
qubit is subjected to DC noise (Fig. \ref{fig:BF1-Pauli2} (c)), the
fidelity can be further maintained, but the variation is quite different
from Fig. \ref{fig:BF1-Pauli2} (a).

It should be noted that when the first (second) qubit is subjected
to BF (AD) noise contour variation of the fidelity can be obtained
by interchanging the $X$ and $Y$ axes in Fig. \ref{fig:AD1-Pauli2}
(a). Due to this reason, these cases will not be further discussed.
Similarly, Fig. \ref{fig:BF1-Pauli2} (a) also shows the variation
of fidelity when both qubits travel through the PF channels.

Another interesting possibility would be to consider the first and
second qubits evolving under PF and DC noise, respectively. It can
be observed from Fig. \ref{fig:PF-DC} (a) that DC noise affects the
fidelity comparatively less than other noises. Finally, we consider
both the qubits subjected to DC noise in Fig. \ref{fig:PF-DC} (b).
The symmetric decay in fidelity shown in this case has a specific
characteristic that even the lowest value of the obtained fidelity
is an appreciable amount when compared with the remaining cases.

\begin{figure}
\centering{}\includegraphics[angle=0,scale=0.4]{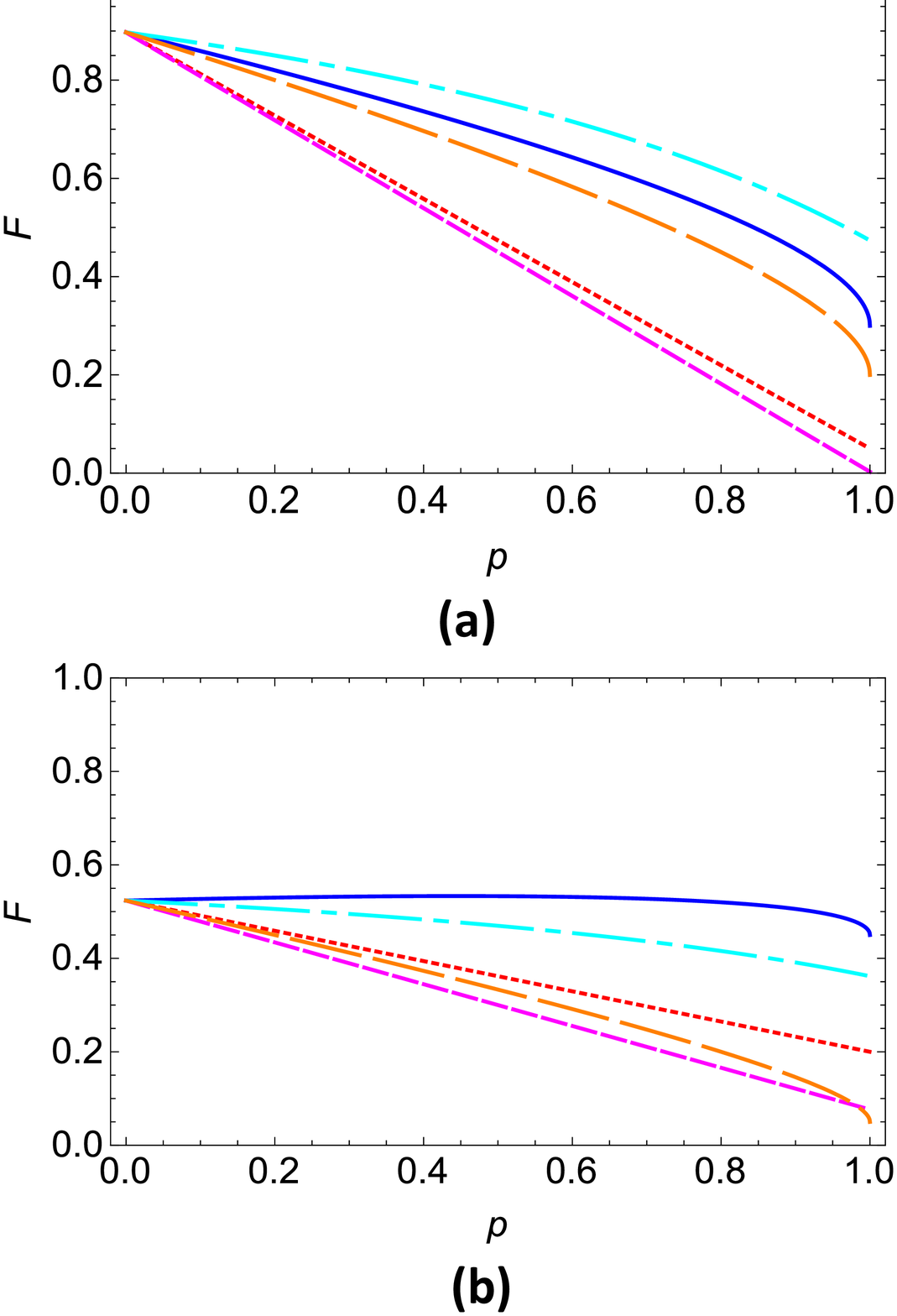}\protect\protect\protect\caption{\label{fig:2DAD}The effect of noise on the fidelity obtained in the
QPC protocol when the first qubit is subjected to AD noise, while
the second qubit is subjected to different noisy channels. The smooth
(blue), large-dashed (orange) lines correspond to the effect of AD
noise on the second qubit for the initial Bell state $|\psi^{\pm}\rangle$
and $|\phi^{\pm}\rangle$, respectively. Similarly, the dotted (red),
small-dashed (magenta), large-dot dashed (cyan) curves correspond
to the fidelity of the obtained states when Bob's qubit is subjected
to BF, PF and DC noises, respectively. In (a) and (b), the decoherence
rate for the Alice's channel is 0.2 and 0.8, respectively. In the
plots, (also in the following figures) the probability of error $p_{2}$
is written as $p$. }
\end{figure}

It is also important to observe a comparative analysis of the effect
of different kinds of noisy channels. Specifically, if we consider
the first qubit travels through a specific channel with the known
rate of error than the effect of noisy environment due to the second
qubit is studied here in two dimensional variation. Specifically,
in Fig. \ref{fig:2DAD}, we consider two cases (namely, decoherence
rate $p_{1}=0.2$ and 0.8) of variation of fidelity with/in error
rate for the second channel. The effects of different values of $p_{1}$
are shown in the figure, as for its higher value, the fidelity remains
the same for initial $|\psi^{\pm}\rangle$ state, while falls sharply
for initial $|\phi^{\pm}\rangle$. It can be seen that for a higher
rate of damping the effect of noise on the other qubit is hardly relevant
(cf. Fig. \ref{fig:2DAD} (b)).

\begin{figure}
\centering{}\includegraphics[angle=0,scale=0.4]{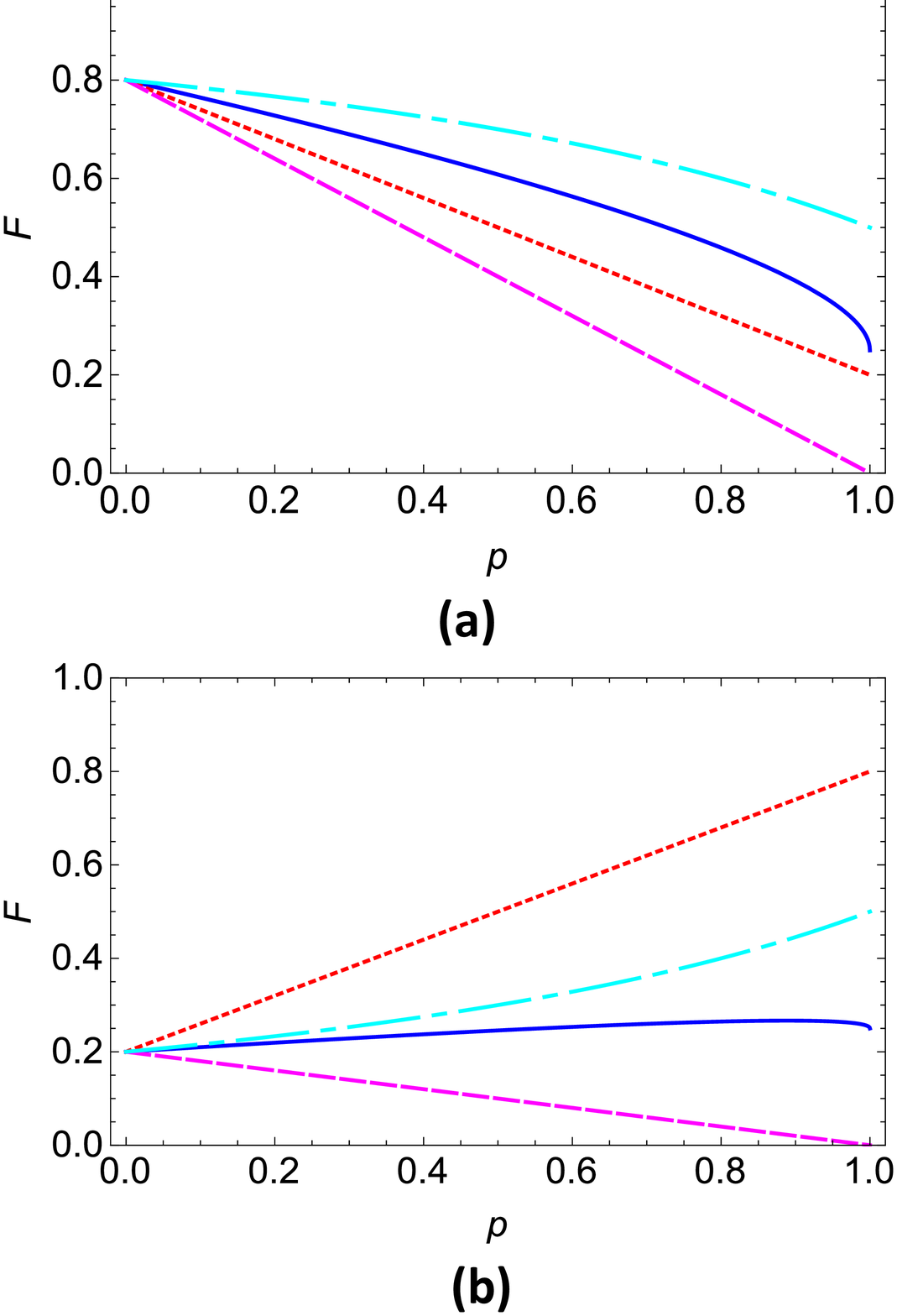}\protect\protect\protect\caption{\label{fig:2DBF}The effect of noise on the fidelity obtained in the
QPC protocol when the first qubit is subjected to BF noise, while
the second qubit is subjected to different noisy channels. The smooth
(blue), dotted (red), dashed (magenta), dot dashed (cyan) curves correspond
to the fidelity of the obtained states when Bob's qubit is subjected
to AD, BF, PF and DC noises, respectively. In (a) and (b), the probability
of error in Alice's channel is 0.2 and 0.8, respectively. }
\end{figure}

Similarly, when the first qubit evolves under the effect of BF channel
with low error rate, the second qubit subjected to DC noise performed
best as shown in Fig. \ref{fig:2DBF} (a). However, in Fig. \ref{fig:2DBF}
(b) the second qubit under BF channel outperformed it for higher probability
of error. Interestingly, the two dimensional cut shown in Fig. \ref{fig:2DBF}
(b) the fidelity is seen to increase with increasing probabilities
of errors in the second channel.

\begin{figure}
\centering{}\includegraphics[angle=0,scale=0.4]{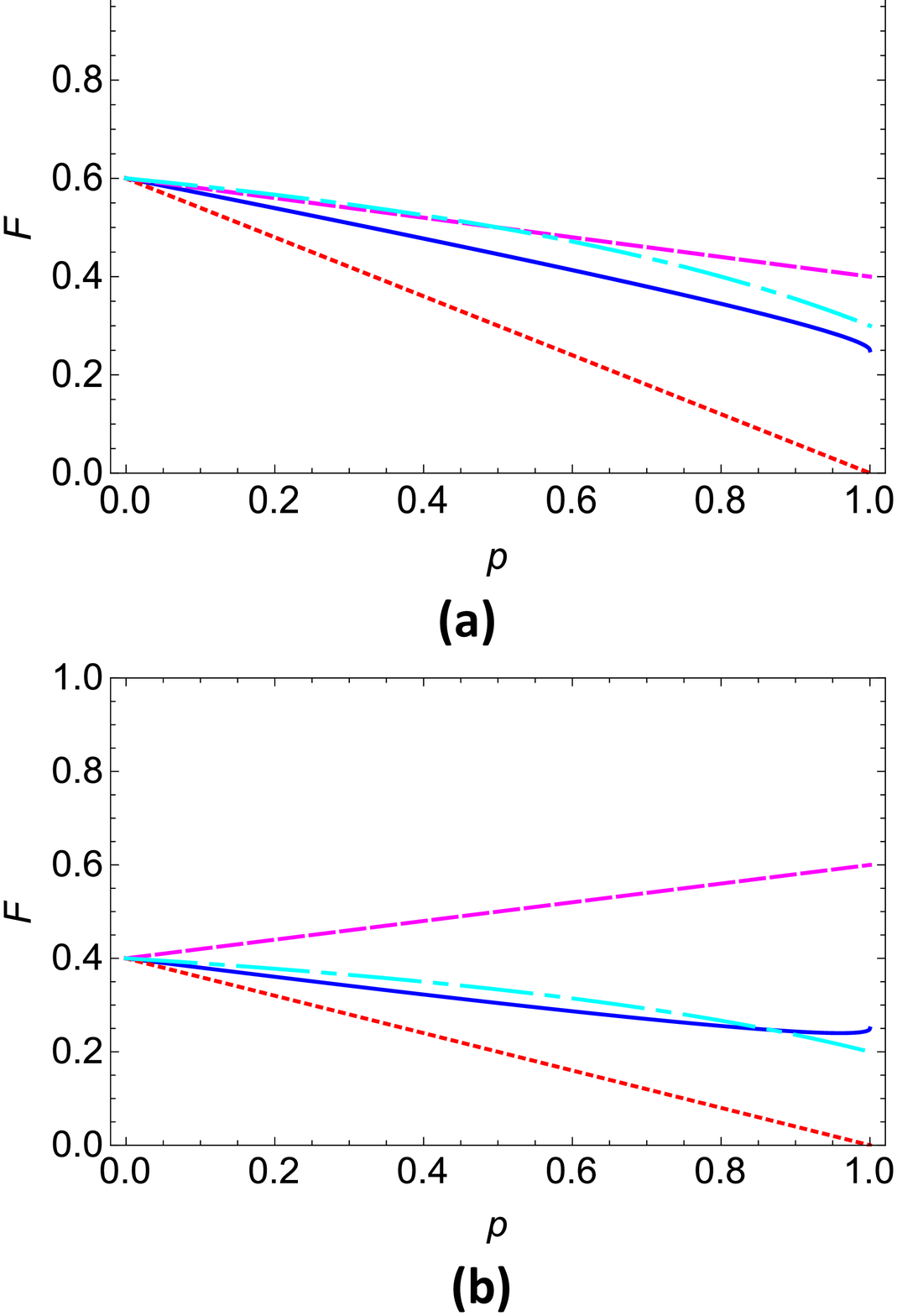} \protect\protect\protect\caption{\label{fig:2DPF}The effect of noise on the fidelity obtained in the
QPC protocol when the first qubit is subjected to PF noise, while
the second qubit is subjected to different noisy channels. The smooth
(blue), dotted (red), dashed (magenta), dot dashed (cyan) curves correspond
to the fidelity of the obtained states when Bob's qubit is subjected
to AD, BF, PF and DC noises, respectively. In (a) and (b), the probability
of error in Alice's channel is 0.4 and 0.6, respectively. }
\end{figure}

When the first qubit is fixed to be traveling through a PF channel
with intermediate error rate, the second qubit evolving under DC noise
suffers most (cf. Fig. \ref{fig:2DPF}). As the fidelity in this case
is initially highest for lower values of error in the second channel,
becoming second to the PF channel for higher errors. The obtained
fidelity in the same case, i.e., second qubit under DC noise, falls
comprehensively even below AD noise for sufficiently large values
of error rates for the second channel.

\begin{figure}
\centering{}\includegraphics[angle=0,scale=0.4]{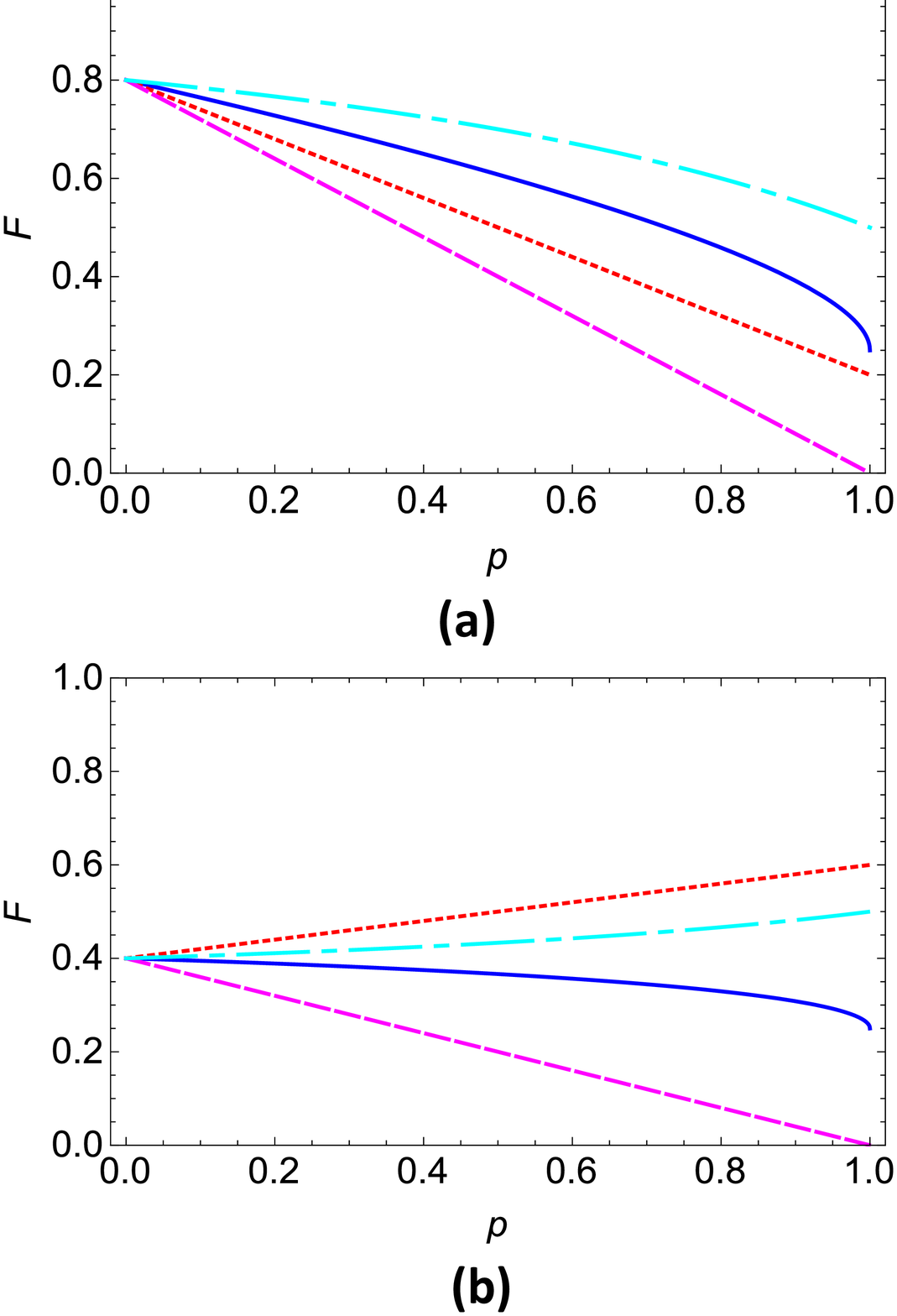}\protect\protect\protect\caption{\label{fig:2DDC}The effect of noise on the fidelity obtained in the
QPC protocol when the first qubit is subjected to DC noise, while
the second qubit is subjected to different noisy channels. The smooth
(blue), dotted (red), dashed (magenta), dot dashed (cyan) curves correspond
to the fidelity of the obtained states when Bob's qubit is subjected
to AD, BF, PF and DC noises, respectively. In (a) and (b), the probability
of error in Alice's channel is 0.2 and 0.6, respectively. }
\end{figure}

Finally, we will consider when the first qubit subjected to DC noise
in Fig. \ref{fig:2DDC}. Here, we can observe that for higher values
of error in the first channel though fidelity in two of the cases
appears increasing, but it is consistent with corresponding contour
plots (cf. Figs. \ref{fig:AD}-\ref{fig:PF-DC}). It is worth commenting
here that fidelity obtained when the second qubit is evolving under
the BF channel improves considerably to be better than in damping
and DC noise.

So far, we have been discussing the effect of noise on the qubits
which are used by Alice and Bob to obtain random key strings. However,
it is also desirable that during eavesdropping checking stage the
presence of Eve and noise can be discriminated. This was the motivation
of our last work \cite{RDshrama2015}, where we have reported that
different suitable choices of decoy qubits and eavesdropping checking
subroutines suitable over different noisy channels. As the GV subroutine
was explicitly discussed in Ref. \cite{RDshrama2015}, we will only
discuss the effect of noise on the semi-quantum subroutine.

\subsection{Effect of noise on semi-quantum subroutine}

It is worth noting that unlike BB84 or GV subroutine, where the principle
of security varies from each other, in a semi-quantum subroutine,
the security can be inherently achieved by using one of them. Specifically,
in a semi-quantum subroutine, a quantum party prepares the decoy qubits
and classical party reflects them to quantum party, who later calculates
the error rate for that. Depending upon the choice of decoy qubits
BB84 or GV subroutine can be used.

However, from the point of view of the effect of noise on this subroutine,
to and fro communication of decoy qubits should be considered while
studying the effect of noisy environments. Here, for the simplicity
of discussion, we will consider the same kind of noise with the same
error rate acting on certain qubit while its to and fro travel.

\begin{figure}
\centering{}\includegraphics[scale=0.55]{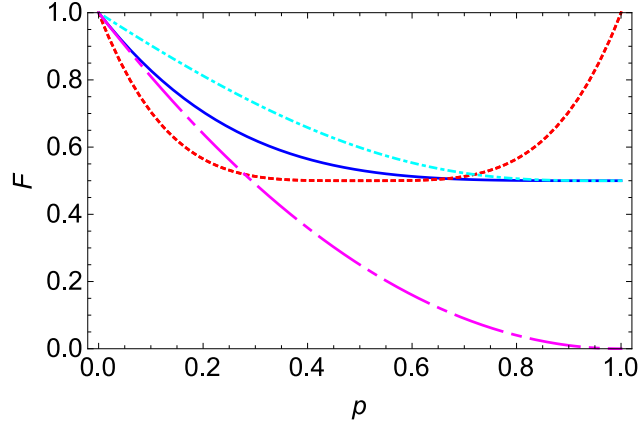}\protect\protect\protect\caption{\label{fig:semi-quantum}The effect of noise on the fidelity of the
quantum states used by TP for eavesdropping checking in the SQPC protocol
when the first both the qubits are subjected to AD noise for initial
choice of Bell states as $|\psi^{\pm}\rangle$ and $|\phi^{\pm}\rangle$
in the smooth (blue) and dashed (magenta) lines, respectively. The
dotted (red) and dot dashed (cyan) curves correspond to the fidelity
of the obtained states when both qubits are affected by BF (equivalently
PF) and DC noises, respectively.}
\end{figure}

Specifically, when both the qubits undergo damping during both rounds
of travels, the obtained fidelity expressions are

\begin{equation}
\begin{array}{lcl}
F & = & \frac{1}{4}\left(\left(-2+p_{2}\right){}^{2}-2p_{1}\left(-2+p_{2}\right)\left(-1+2p_{2}\right)+p_{1}^{2}\left(1+2\left(-2+p_{2}\right)p_{2}\right)\right)\end{array}\label{eq:ADP1-ADP2}
\end{equation}
and

\begin{equation}
F=\frac{1}{4}\left(-2+p_{1}+p_{2}\right){}^{2}\label{eq:ADP1-ADP2 PHIPLUS}
\end{equation}
for the initial choice of Bell state as $|\psi^{\pm}\rangle$ and
$|\phi^{\pm}\rangle$, respectively. In the remaining cases, when
the first qubit is subjected to damping and the second qubit is considered
under BF, PF, and DC noises, the obtained fidelity is calculated as

\begin{equation}
F=\frac{1}{4}\left(-2+p_{1}\right)\left(-2+p_{1}\left(1-2p_{2}\right){}^{2}-4\left(-1+p_{2}\right)p_{2}\right),\label{eq:ADP1-BFP2}
\end{equation}

\begin{equation}
F=\frac{1}{4}\left(\left(-2+p_{1}\right){}^{2}+8\left(-1+p_{1}\right)p_{2}-8\left(-1+p_{1}\right)p_{2}^{2}\right),\label{eq:ADP1-PFP2}
\end{equation}
and

\begin{equation}
F=\frac{\left(-2+p_{1}\right)\left(-8+4p_{1}\left(-1+p_{2}\right){}^{2}+\left(12-5p_{2}\right)p_{2}\right)}{4\left(-2+p_{2}\right){}^{2}},\label{eq:ADP1-DNP2}
\end{equation}
respectively. In another case, when the first qubit is subjected to
BF channel and second qubit evolves under the effect of BF, PF, and
DC noise the analytic expressions of fidelity are

\begin{equation}
F=1-2p_{1}\left(1-2p_{2}\right){}^{2}+2p_{1}^{2}\left(1-2p_{2}\right){}^{2}+2\left(-1+p_{2}\right)p_{2},\label{eq:BFP1-BFP2}
\end{equation}

\begin{equation}
F=-\frac{\left(1+2\left(-1+p_{1}\right)p_{1}\right)\left(1+2\left(-1+p_{2}\right)p_{2}\right)}{-1+\left(-1+p_{1}\right){}^{2}\left(-1+p_{2}\right)p_{2}},\label{eq:BFP1-PFP2}
\end{equation}
and

\begin{equation}
\begin{array}{lcl}
F & = & \frac{1}{2\left(-2+p_{2}\right){}^{2}}\left[8-16p_{1}\left(-1+p_{2}\right){}^{2}+16p_{1}^{2}\left(-1+p_{2}\right){}^{2}+p_{2}\left(-12+5p_{2}\right)\right],\end{array}\label{eq:BFP1-DNP2}
\end{equation}
respectively. Similarly, when the first (second) qubit is affected
by PF (DC) noise, then the fidelity can be calculated as

\begin{equation}
F=\frac{\left(1+2\left(-1+p_{1}\right)p_{1}\right)\left(8+p_{2}\left(-12+5p_{2}\right)\right)}{2\left(-2+p_{2}\right){}^{2}}.\label{eq:PFP1-DNP2}
\end{equation}
Finally, when both the qubits are subjected to DC noise, we obtain
the expression for fidelity as

\begin{equation}
\begin{array}{lcl}
F & = & \frac{1}{2\left(-2+p_{1}\right){}^{2}\left(-2+p_{2}\right){}^{2}}\left[4\left(8+p_{2}\left(-12+5p_{2}\right)\right)-4p_{1}\left(12+p_{2}\left(-20+9p_{2}\right)\right)+p_{1}^{2}\left(20+p_{2}\left(-36+17p_{2}\right)\right)\right].\end{array}\label{eq:DNP1-DNP2}
\end{equation}
The fidelity expressions for the remaining cases are not explicitly
written here, as those expressions can be derived from the above expressions
by symmetry argument. The analysis of the obtained fidelity expression
is simplified here by considering the case when both the qubits are
affected by the same kind of noise with the same error rate. In this
particular case, as shown in Fig. \ref{fig:semi-quantum}, it is observed
that the fidelity for the damping effect, for $|\phi^{\pm}\rangle$
as the initial choice of TP, gradually decays to zero. Further, for
the state evolving over the BF or PF channel, a revival of with an
increasing error rate is observed. For the remaining cases (i.e.,
for $|\psi^{\pm}\rangle$ state under AD channels and an arbitrary
Bell state subjected to DC noise), fidelity falls gradually and becomes
constant around $p=0.5$.

\section{Conclusion\label{sec:Conclusion}}

Two protocols for QPC have been proposed. The essential beauty of
the present work underlies in the fact that both of the proposed protocols
for QPC are fundamentally different from all the existing protocols
of QPC. Specifically, the first protocol is designed solely using
orthogonal states and the security of the protocol does not rely on
conjugate coding. This is interesting because of the fact that BB84
protocol \cite{bb84} for QKD and early protocols of secure quantum
communication (see Chapter 8 of Ref. \cite{book} for a review), whose
security arise from the fact that one cannot perform simultaneous
measurements in mutually unbiased bases (conjugate coding), led to
a perception that unconditional security of QKD and other quantum
tasks arises from the conjugate coding (i.e., through the use of non-orthogonal
states as the states prepared in mutually unbiased bases are not orthogonal
to each other). Interestingly, in Refs. \cite{vaidman-goldenberg,N09}
it has been firmly established that QKD is possible without using
non-orthogonal states, i.e., by using orthogonal states only. Subsequently,
a set of orthogonal-state-based protocols for various secure quantum
communication tasks has been proposed by some of the present authors
\cite{book,With-preeti,dsqc-ent-swap,QKA-cs}, and those protocols
have established that most of the quantum cryptographic tasks that
can be performed using non-orthogonal states can also be performed
using orthogonal states. Specifically, it has been established that
one can perform quantum secure direct communication \cite{book,With-preeti,dsqc-ent-swap},
QKA \cite{QKA-cs}, etc., solely using orthogonal states. These orthogonal-state-based
protocols are fundamentally different from the conjugate coding based
schemes as their security does not arise from our inability to perform
simultaneous measurement using two mutually unbiased bases. However,
until now, no effort has been made to design an orthogonal-state-based
protocol for QPC. Thus, the first protocol of the present letter is
fundamentally different from all the existing protocols for QPC, and
it is unique in that sense. Interestingly, the second protocol proposed
here is also an orthogonal-state-based, but in the case of the second
protocol we did not stress much on this characteristic, because it
has another unique characteristic, which has not been explored in
any of the existing schemes for QPC: the protocol is semi-quantum
in nature, i.e., it allows some of the users to be classical. These
two protocols answer couple of questions (in context of QPC) that
are foundationally important. For example, both the protocol answer:
Which quantum properties are essential for the implementation of schemes
of QPC? It establishes that conjugate coding is not required and ensuring
simultaneous non-availability of all the pieces of information to
Eve after splitting it into several pieces is sufficient to ensure
security of schemes for QPC. Specifically, conjugate-coding-based
schemes achieve security by hiding the basis information of two mutually
unbiased bases, whereas an orthogonal-state-based employs geographical
information splitting using PoP technique. The second protocol answers
another question: How much quantumness is needed (or in other words,
how many quantum users are needed) to implement a scheme for QPC?
It is established that only TP needs to be quantum and all other users
can be classical. This is in sharp contrast to the existing protocols
where all the users are required to be quantum. Thus, the second protocol
of ours clearly uses reduced quantum resources, as two parties comparing
their assets are completely classical in nature. However, there exists
a trade-off between the amount of quantum resources used and the qubit
efficiency achieved. Specifically, the qubit efficiency of the SQPC
protocol (i.e., our second protocol) is found to be much lower compared
to the first protocol where all the users are quantum in nature. Clearly,
this happens because in SQPC, a lesser number of parties possess quantum
resources and that increases the requirement of $q$ and $b$ for
accomplishing the same task (i.e., to communicate the same amount
of classical information $c$ by following the restrictions of the
same cryptographic task).

The feasibility of both of the proposed schemes is analyzed under
well-known noise models, such as AD, BF, PF, and DC. The study has
led to many interesting conclusions. In general, the effect of any
specific noise model is independent of the choice of the initial Bell
state. However, it is observed that when both the qubits are subjected
to AD noise, the fidelity expressions depend on the parity of the
Bell state. Further, the study has established that the quantum state
is least affected due to DC illustrated through a higher value of
fidelity. Interestingly, it has also been observed that the higher
error rates in one of the quantum channel can also lead to positive
effects. Specifically, a few cases have been observed where the higher
decoherence rate in one of the quantum channel had resulted in higher
fidelity compared to the situation having lower error rates in the
same channel. Another interesting result has been obtained when both
the qubits were traveling over a BF channel, as in this case a revival
in fidelity with increasing error rate in a quantum channel has been
observed. The effect of noise in the eavesdropping checking for SQPC
protocol is also considered, as it is necessary to differentiate between
noise and Eve. It has been observed that the parity 1 Bell states
have least fidelity when subjected to AD noise.

Keeping the above in mind, we conclude the letter by noting that a
QPC scheme neither requires user other than TP to be quantum, nor
it requires to use the quantum states prepared in mutually unbiased
bases. However, to implement a scheme for QPC in a realistic situation,
characterization of the channel (knowledge of noise(s) present in
the quantum channel) would play a very important role, as in the absence
of this knowledge, TP may reach to an incorrect conclusion.

\textbf{Acknowledgment: }AP and KT thank Defense Research \& Development
Organization (DRDO), India for the support provided through the project
number ERIP/ER/1403163/M/01/1603.

\end{document}